\documentclass[12pt, centerh1]{article}
\usepackage{natbib}
\usepackage[margin=1in]{geometry}
\usepackage{amsmath,natbib,multirow}
\usepackage{amsfonts,mathrsfs,moreverb,lipsum}
\usepackage{graphicx,colonequals}
\usepackage{fixltx2e}
\usepackage{color}
\usepackage{caption}
\usepackage{mathtools}
\usepackage{pbox}
\usepackage{subfig}
\newcommand{\btheta}{{\boldsymbol{\theta}}}
\newcommand{\bvtheta}{{\boldsymbol{\vartheta}}}
\newcommand{\pig}{\pi_g}

\newcommand{\bmu}{{\boldsymbol{\mu}}}
\newcommand{\bSig}{{\boldsymbol{\Sigma}}}

\newcommand{\vecmu}{\mbox{\boldmath$\mu$}}

\newcommand{\vecx}{\mathbf{x}}

\newcommand{\vecX}{\mathbf{X}}

\newcommand{\vecV}{\mathbf{V}}

\newcommand{\loada}{\mathbf\Lambda}

\newcommand{\matsig}{\mathbf\Sigma}

\newcommand{\vecC}{\mathbf{C}}

\newcommand{\vecalp}{\boldsymbol{\alpha}}

\title{Skewed Distributions or Transformations? Modelling Skewness for a Cluster Analysis}
\author{Michael P.B. Gallaugher$^*$, Paul D. McNicholas$^*$\\ Volodymyr Melnykov$^{**}$, Xuwen Zhu$^{**}$ }
\date{\small $^*$ Dept.\ of Mathematics \& Statistics, McMaster University, Hamilton, Ontario, Canada. \\ $^{**}$ Department of Information Systems, Statistics \& Management Science, University of Alabama, Tuscaloosa, Alabama, United States}
\begin{document}
\maketitle{}

\begin{abstract}
Because of its mathematical tractability, the Gaussian mixture model holds a special place in the literature for clustering and classification. For all its benefits, however, the Gaussian mixture model poses problems when the data is skewed or contains outliers. Because of this, methods have been developed over the years for handling skewed data, and fall into two general categories. The first is to consider a mixture of more flexible skewed distributions, and the second is based on incorporating a transformation to near normality. Although these methods have been compared in their respective papers, there has yet to be a detailed comparison to determine when one method might be more suitable than the other. Herein, we provide a detailed comparison on many benchmarking datasets, as well as describe a novel method to assess cluster separation.
\\[-10pt]

\noindent\textbf{Keywords}: mixture models; skewed distributions; transformations; cluster overlap.
\end{abstract}

\section{Introduction}
Clustering and classification look at finding and analyzing underlying group structures in data. %There is no one agreed upon definition for a cluster and a detailed discussion of this can be found in \cite{mcnicholas16a,mcnicholas16b}, who follow others in advocating a .
One common method used for clustering is model-based, and generally makes use of a $G$-component finite mixture model. A random variable $\vecX$ from a finite mixture model has density
$$
f(\vecx~|~\bvtheta)=\sum_{g=1}^G\pig f_g(\vecx~|~\btheta_g),
$$
where $\bvtheta=\left(\pi_1,\pi_2,\ldots,\pi_G,\btheta_1,\btheta_2,\ldots,\btheta_G\right)$, $f_g(\cdot)$ is the $g$th component density, and $\pig>0$ is the $g$th mixing proportion such that $\sum_{i=1}^G\pig=1$.

The association between clustering and mixture models can be traced all the way back to \cite{tiedeman55}, as discussed in \cite{mcnicholas16a}, and the earliest use of a mixture model for clustering, specifically a Gaussian mixture model, can be found in \cite{wolfe65}. \cite{baum70} and \cite{scott71} are other examples of early work in this area, and a recent review of model-based clustering is given by \cite{mcnicholas16b}.

Due to its mathematical tractability, the Gaussian mixture model holds a special place in the clustering literature. For all its benefits, however, the Gaussian mixture model poses problems when dealing with data that is either skewed or contains outliers. Specifically, in the presence of cluster skewness and/or outliers, the Gaussian mixture tends to over fit the number of groups. Therefore, many methods have been proposed over the years to alleviate this issue; however, the main streamline focuses on developing mixtures with components capable of modelling skewness and/or kurtosis. It can be achieved by either choosing some flexible traditional distributions or making use of transformations. The latter approach assumes that upon some transformation, the data become nearly normal. Then, a component-wise back transformation of the Gaussian mixture leads to the so-called transformation mixture model. Although these methods have been compared to a certain extent in their respective papers, there is still uncertainty as to when one method might be preferred over another. Herein we aim to fill this gap by performing an extensive study on well-known clustering datasets with an extensive set of initialization partitions. In addition to this extensive comparison, a new method for determining cluster separation is also proposed.

%The remainder of this paper is laid out as follows. In Section ~2, a detailed discussion of the underlying methodology is given, as well as some numerical considerations. Section ~3 describes in detail the methods used for comparison, Section ~4 considers the analysis of many real datasets, and we end with a discussion.
\section{Background}
\subsection{Mixtures of Skewed Distributions}
The first class of methods for dealing with skewness is to consider mixtures of more flexible distributions that incorporate skewness, and many such distributions could be considered. Such examples include the skew-$t$ \citep{lin10,vrbik12,vrbik14,lee14,murray14b,murray14a}, the normal-inverse Gaussian (NIG) distribution \citep{karlis09,ohagan16}, the shifted asymmetric Laplace (SAL) distribution \citep{morris13b,franczak14}, the variance-gamma distribution \citep{smcnicholas17}, the generalized hyperbolic distribution \citep{browne15}, the hidden truncation hyperbolic distribution \citep{murray17b}, the joint generalized hyperbolic distribution \citep{tang18}, and the coalesced generalized hyperbolic distribution \citep{tortora19}. For the purpose of this analysis, we will focus on two representative distributions arising from a normal variance-mean mixture model. %The main reason for this choice is the computational efficiency when compared to hidden truncation methods.  
A normal variance-mean mixture model assumes that a random vector $\vecX$ can be written 
$$
\vecX=\vecmu+W\vecalp+\sqrt{W}\vecV,
$$
where $\vecmu$ and $\vecalp$ are the location and skewness, respectively, $W$ is a positive random variable, and $\vecV\sim\mathcal{N}({\bf 0},\matsig)$.

If $W\sim\text{Gamma}(\gamma,\gamma)$, then the result is the variance-gamma distribution \citep{mcnicholas17} and its density is
\begin{align*}
f_{\text{VG}}(\vecx|\bvtheta)=&\frac{2\gamma^{\gamma}\exp\left\{(\vecx-\vecmu)'\matsig^{-1}\vecalp)\right\}}{(2\pi)^{\frac{p}{2}}| \matsig |^{\frac{1}{2}}\Gamma(\gamma)}  \left(\frac{\delta(\vecx;\vecmu,\matsig)}{\rho(\vecalp,\matsig)+2\gamma}\right)^{\frac{\left(\gamma-{p}/{2}\right)}{2}} \\
&\qquad\qquad\times  K_{\left(\gamma-\frac{p}{2}\right)}\left(\sqrt{\left[\rho(\vecalp,\matsig)+2\gamma\right]\left[\delta(\vecx;\vecmu,\matsig)\right]}\right),
\end{align*}
where $\delta(\vecX;\vecmu,\matsig)=(\vecx-\vecmu)'\matsig^{-1}(\vecx-\vecmu)$, $\rho(\vecalp,\matsig)=\vecalp'\matsig^{-1}\vecalp$,  $\gamma\in\mathbb{R}^+$ is a concentration parameter, and $K_{\lambda}(x)$ is the modified Bessel function of the third kind with index $\lambda$. 
Likewise, if $W\sim \text{GIG}(\omega,1,\lambda)$, where GIG represents the generalized inverse Gaussian distribution with the parameterization used by \cite{browne15}, then the result is the generalized hyperbolic distribution, with density 
\begin{align*}
f_{\text{GH}}(\vecx|\bvtheta)=&\frac{\exp\left\{(\vecx-\vecmu)\matsig^{-1}\vecalp' \right\}}{(2\pi)^{\frac{np}{2}}| \matsig |^{\frac{1}{2}} K_{\lambda}(\omega)}  \left(\frac{\delta(\vecx;\vecmu,\matsig)+\omega}{\rho(\vecalp,\matsig)+\omega}\right)^{\frac{\left(\lambda-\frac{p}{2}\right)}{2}} \\ & \qquad\qquad\times
 K_{\left(\lambda-{p}/{2}\right)}\left(\sqrt{\left[\rho(\vecalp,\matsig)+\omega\right]\left[\delta(\vecx;\vecmu,\matsig)+\omega\right]}\right),
\end{align*}
where $\lambda\in \mathbb{R}$ is the index parameter and $\omega\in\mathbb{R}^+$ is a concentration parameter.

Likewise, the skew-t and NIG distributions can be derived in a similar manner. Parameter estimation is performed using an expectation conditional maximization (ECM) algorithm, and the details for the VG and GH distributions can be found in \cite{mcnicholas17} and \cite{browne15}, respectively.

\subsection{Infinite Likelihood Problem}
One numerical aspect of the variance-gamma distribution that must be considered is the infinite likelihood problem. This was discussed by \cite{franczak14} for the skewed asymmetric Laplace (SAL) distribution, and occurs when $\hat{\vecmu}_g\rightarrow\vecx_i$. As the SAL distribution is a special case of the variance-gamma with $\gamma=1$, it is not surprising that this also occurs for the variance-gamma distribution. This is due to the density being unbounded when $\hat{\vecmu}_g\rightarrow\vecx_i$ and $\gamma<p/2$, as discussed in \cite{Nit15}. This occurs because the Bessel function goes to infinity as the argument approaches 0, and 
$$
 \left(\frac{\delta(\vecx;\vecmu,\matsig)}{\rho(\vecalp,\matsig)+2\gamma}\right)^{\frac{\left(\gamma-{p}/{2}\right)}{2}}\rightarrow \infty
$$
if $\vecmu\rightarrow\vecx_i$ and $\gamma<p/2$.

The solution to this problem is not trivial. In \cite{franczak14}, the authors propose running the ECM algorithm to the point where this occurs, go back one iteration and set $\hat{\vecmu}_g$ to be the value at the preceding iteration, and update the skewness accordingly. An alternative solution is to bound the density in this situation \citep{Nit15}. One problem with both of these solutions is that, at that point in the algorithm, the parameter estimates have most likely entered an unstable part of the parameter space. 

Therefore, with all of this taken into consideration, we propose restricting $\gamma$ so that if $\hat{\gamma}_g<p/2$ then we set $\hat{\gamma}_g=p/2$. 

It is important to note that this scenario does not occur for the generalized hyperbolic, skew-$t$ and NIG distributions because, in all of these cases, the $\delta(\cdot)$ term is accompanied by a positive value $\omega$ thus ensuring boundedness of the density function for all values of $\lambda$.

\subsection{Transformation Methods}
The second class of methods to handle skewness in a cluster analysis is the use of some transformation to near normality, introduced by \cite{zhu18, melnykovandzhu18, melnykovandzhu18b}.
The basic idea is to assume that there exists a transformation, $\mathcal{T}(\vecX~|~\loada)$, where $\underline{\vecx}=(\vecx_1,\vecx_2,\ldots,\vecx_p)$ is the original data vector, and $\loada=(\lambda_1,\lambda_2,\ldots,\lambda_p)'$ is a transformation vector, such that 

$$
\mathcal{T}(\underline{\vecx}~|~\loada)=(\mathcal{T}(\vecx_1|\lambda_1),\mathcal{T}(\vecx_2|\lambda_2),\ldots,\mathcal{T}(\vecx_p|\lambda_p))\sim\mathcal{N}(\vecmu,\matsig),
$$
where $\mathcal{N}(\cdot)$ represents the $p$-variate normal distribution with mean $\vecmu$ and covariance $\matsig$, $\lambda_1,\lambda_2,\ldots,\lambda_p$ are marginal transformation parameters responsible for each one of the $p$ dimensions, respectively. Herein two different univariate transformations are considered, namely the power and Manly transformations.

The power transformation, proposed by \cite{yeo00}, is defined as
$$
\mathcal{T}(x|\lambda)=
\left\{
\begin{array}{ccc}
[(x+1)^{\lambda}-1]/ \lambda&& \text{if \quad} (x\ge 0,\lambda\ne0),\\
\log(x+1) &&\text{if \quad} (x\ge 0,\lambda=0),\\
-[(-x+1)^{2-\lambda}-1]/(2-\lambda) & & \text{if \quad}(x<0, \lambda\ne2),\\
-\log(-x+1) &&\text{if \quad} (x<0, \lambda=2).
\end{array}
\right. 
$$

The Manly transformation (also called the exponential transformation), introduced first by \cite{manly76}, is defined as
$$
\mathcal{T}(x|\lambda)=
\left\{
\begin{array}{ccc}
[\exp\{\lambda x\}-1]/\lambda &&\text{if \quad} \lambda\ne 0,\\
x && \text{otherwise.}\\
\end{array}
\right. 
$$

Both of these transformations have been shown to handle both positive and negative skewness. By applying the back-transformation from $p$-variate normal distribution, the corresponding transformation-based density can be written as
$$
f_{\mathcal{T}}(\vecx~|~\bvtheta)= \phi(\mathcal{T}(\vecx~|~\loada); \bmu, \bSig) J_{\mathcal{T}}(\vecx~|~ \loada),
$$
where $J_{\mathcal{T}}(\vecx~|~ \loada) =  \vert \partial \mathcal{T}(\vecx~|~\loada) / \partial \vecx' \vert$ represents the Jacobian derived based on the back-transformation from normal distribution. 
For the Manly transformation, its Jacobian can be written as $J_{\mathcal{T}}(\vecx~|~ \loada) \equiv \exp\{\loada' \vecx\}$. For the power transformation, $$J_{\mathcal{T}}(\vecx~|~ \loada) \equiv \prod_{j=1}^p (|x_j|+1)^{\text{sgn}(x_j)(\lambda_j-1)},$$
where
$$
\text{sgn}(x)=
\left\{
\begin{array}{ccc}
1 &&\text{if \quad} x>0,\\
0 && \text{if \quad} x=0,\\
-1&& \text{if \quad} x<0.\\
\end{array}
\right. 
$$

\section{Measures Used for Comparison} 
\subsection{Multivariate Skewness and Kurtosis}
For the purposes considered here, an assessment of component skewness and kurtosis is desirable. While it is straightforward to consider the univariate skewness and kurtosis for each dimension, for higher dimensions these characteristics become difficult to assess.  Therefore, the definition of Mardia's multivariate skewness and kurtosis \citep{mardia70} is employed, as this gives a single measure for skewness and kurtosis, and it provides a test for assessing multivariate normality.

The multivariate skewness, $\beta_{1,p}$ for a multivariate random vector $\vecX$ with mean vector $\vecmu=(\mu_1,\mu_2,\ldots,\mu_p)$ and covariance $\matsig$ is defined as
$$
\beta_{1,p}=\sum_{r,s,t}\sum_{r',s',t'}\sigma^{rr'}\sigma^{ss'}\sigma^{tt'}\mu_{111}^{(rst)}\mu_{111}^{(r's't')},
$$
where $\sigma^{ij}$ is the $(i,j)$ element of the inverse covariance matrix $\matsig^{-1}$ and 
$$
\mu_{111}^{rst}=\mathbb{E}[(X_r-\mu_r)(X_s-\mu_s)(X_t-\mu_t)].
$$
Under multivariate normality, $\beta_{1,p}=0$.

The multivariate kurtosis is defined similarly and is given by
$$
\beta_{2,p}=\mathbb{E}[(\vecX-\vecmu)'\matsig^{-1}(\vecX-\vecmu)]^2,
$$
and under multivariate normality has value $p(p+2)$. Therefore, we report the kurtosis as $\hat{\beta}_{2,p}-p(p+2)$, so that negative kurtosis corresponds to lighter tails and positive kurtosis corresponds to heavier tails. In addition to these values, \cite{mardia70} also provides tests to determine if the skewness and kurtosis are significantly different from what is expected under normality. These values, along with their p-values for the tests can be calculated using the {\sf R} package {\tt psych} \citep{psych}.

\subsection{Cluster Overlap}
One property of a dataset we consider for comparing the two classes of methods is cluster separation. One possible measure of the proximity of two components is a so called pairwise overlap \citep{maitra10} defined as the sum of two misclassification probabilities. The larger the sum, the more substantial overlap between the components is observed. An outline of the method we employ is described as follows.
\begin{enumerate}
\item {\bf Density Estimation}:\\
In this step, a density function $\hat{f}_g(\vecx)$ is estimated for each group $g\in\{1,2,\ldots,G\}=\mathcal{G}$ with mixing proportions $\pi_1,\pi_2,\ldots,\pi_G$.
\item {\bf Simulation}:\\
For each group $g$, simulate $N$ observations. Denote these observation matrices by $\vecx_g=(\vecx_{g1},\vecx_{g2},\ldots,\vecx_{gN})$. 
\item {\bf Calculation}:\\
For each $\vecx_g$ and $g\in\mathcal{G}$, let $\vecC_g$ denote an $N$ dimensional vector with entry $i$ being 
$$
\vecC_g\{i\}=\text{argmax}_{h\in\{1,2,\ldots,G\}}\pi_h\hat{f}_g(\vecx_{hi})
$$
\item {\bf Map}:\\
For each $g, h\in\mathcal{G}$, let 
$$
p_{gh}=\frac{1}{N}\sum_{i=1}^N\mathbb{I}(\vecC_g=h)
$$ 
and let the $G\times G$ misclassification map matrix $P$ be defined as $P\{g,h\}=p_{gh}$. 
\end{enumerate}
This general method has been used, for example, in \cite{melnykov16} after fitting the model of interest, and considered pairwise overlap between clusters as the sum of the two misclassification probabilities.

For the purposes considered here, it is desirable that the estimated density captures the true nature of the component and for simulation to be computationally feasible. Herein, two different methods are proposed. The first is to consider an approximation by a mixture of Gaussian distributions,
$$
\hat{f}_g(\vecx|\bvtheta)=\sum_{j=1}^J\pi_j\phi_{p}(\vecmu_j,\matsig_j).
$$

Although not effective for modelling multiple skewed components, a mixture of Gaussian distributions is effective for modelling a single skewed component. Moreover, it is straightforward to fit and then simulate from a Gaussian mixture.
Another method we propose for density estimation is to consider kernel density estimation (KDE). This assumes that the estimate of the density can be written
$$
\hat{f}_g(\vecx)=\frac{1}{n}\sum_{i=1}^nK_{{\bf H}}(\vecx-\vecx_i),
$$
where $n$ is the sample size, ${\bf H}$ is a smoothing matrix, and $K(\cdot)$ is here the Kernel function chosen to be Gaussian in this case. 

The choice of a smoother matrix is not trivial, especially in the multivariate case, and many such matrices have been proposed. However, for our purposes, we consider a diagonal smoother matrix with elements 
$$
h_j=\left(\frac{4}{p+2}\right)^{1/(p+4)}n^{-1/(p+4)}\hat{\sigma}_j,
$$
where $\hat{\sigma}_j$ is the estimated standard deviation of variable $j$ \citep{hardle97}.
\subsubsection{Example: Iris Dataset}
\begin{figure}[!htb]
\centering
\includegraphics[width=0.8\textwidth]{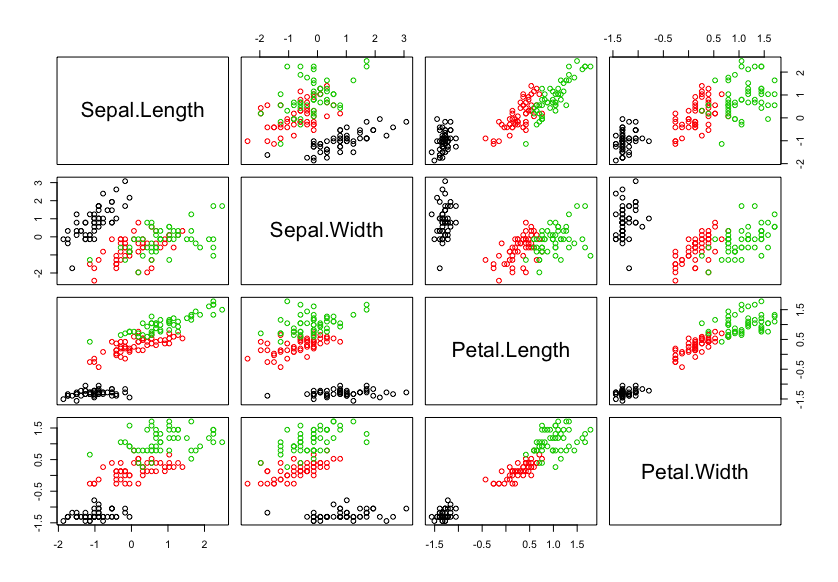}
\caption{Pairs plot of the iris dataset.}
\label{fig:iris}
\end{figure}

\begin{figure}[!htb]%
    \centering
    \subfloat[]{{\includegraphics[width=0.7\textwidth]{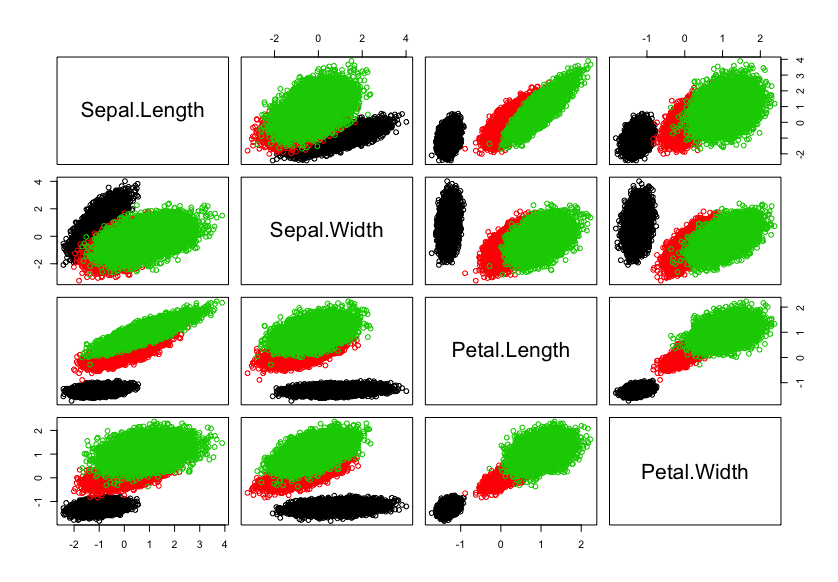} }}%
    \qquad
    \subfloat[]{{\includegraphics[width=0.7\textwidth]{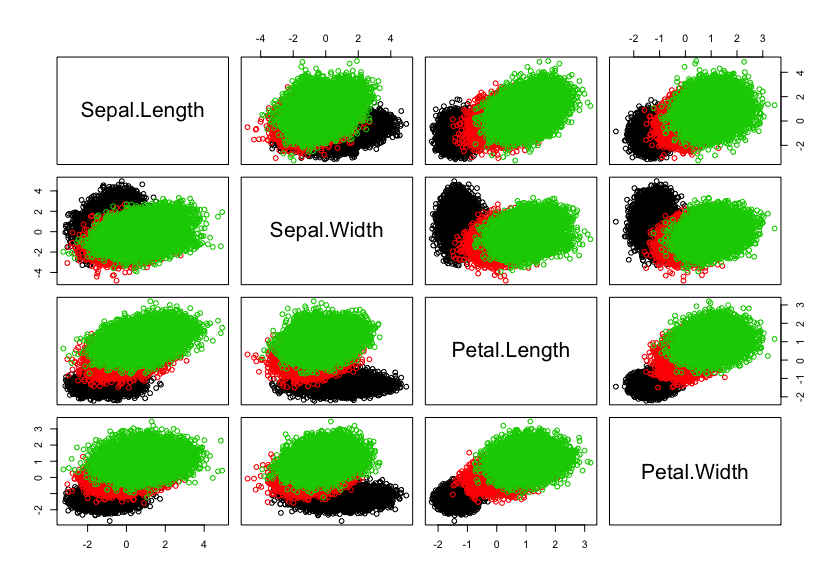} }}%
    \caption{1000 simulated points for each component of the iris dataset simulated using (a)~a fitted mixture of Gaussian distributions and (b) using KDE.}%
    \label{fig:iris2}%
\end{figure}

An example of the cluster overlap procedure described above is now presented on the well known Iris dataset \citep{anderson35}. This dataset provides four measurements on three different species of iris. Figure \ref{fig:iris} displays a pairs plot for the original dataset, and Figure~\ref{fig:iris2} shows 1000 simulated points for each cluster using the two different density estimation methods described above. Table \ref{tab:MapIris} shows the resultant misclassification maps for the two different density estimation methods.

\begin{table}
\centering
\caption{Gaussian mixture and KDE misclassification maps for the iris dataset.}
\begin{tabular}{cc}
\hline
Gaussian Mixture Map:
$\left(\begin{tabular}{rrr}
1.00 & 0.00 & 0.00 \\ 
  0.00 & 0.98 & 0.02 \\ 
  0.00 & 0.03 & 0.97 \\ 
\end{tabular}\right)$&
KDE Map: $\left(\begin{tabular}{rrr}
1.00 & 0.00 & 0.00 \\ 
  0.03 & 0.77 & 0.21 \\ 
  0.00 & 0.17 & 0.83 \\ 
\end{tabular}\right)$\\
\hline
\end{tabular}
\label{tab:MapIris}
\end{table}

It is interesting to note that when using KDE, there is more overlap than when using a Gaussian mixture. Perhaps, it can be explained by the fact that when using a mixture of Gaussian distributions for each component, naturally more points will be simulated closer to the centre of each component in the mixture. In the case of KDE estimation, the points in the tail of the distribution, more importantly the points which lie on the border of the two clusters, have the same probability of being chosen as the points in the centre. 
\clearpage
\subsubsection{General Applicability}
It is important to note that in the scenario of a true cluster analysis this method for measuring cluster overlap is not applicable because the true group labels are not known. However, this is useful for the purposes presented here as we wish to determine if cluster separation does result in different performance for the two different methods. Moreover, this overlap procedure would be useful in other scenarios. One example would be in the case of a discriminant analysis, where the true labels are known. Another would be in the case determining separation of classes of categorical variables such as gender or race in regardless of the method of analysis.

\subsection{Initialization and Convergence}\label{Init}
In order to increase the chances of obtaining the maximum likelihood, many different starting values are considered and then the algorithm is ran to full convergence. Specifically, up to eleven k-means initializations, 1000 soft partitions, up to 100$G$ (where $G$ is the number of groups) unique hard initializations by running 1 iteration of the k-means algorithm, see \cite{melnykov12} for details, and a hierarchical partition using Ward's linkage \citep{ward63}. Note that only 100$G$ hard initializations are considered because the number of unique hard partitions increases with $G$. In addition, although not applicable in practice, but useful for comparison purposes, initializing with the true labels is considered. The final results are obtained by taking the largest likelihood over all these initializations, with the exception of the true labels.

The same convergence criterion was used for both methods. Specifically, the EM algorithm is terminated when $(\ell^{(t+1)}-\ell^{(t)})/|\ell^{(t+1)}|<0.0001$, where $\ell^{(t)}$ is the estimated log-likelihood at iteration $t$. 

%\section{Comparison}
%\subsection{Some Technical Differences}
%One considerable difference between the use of traditional and transformation-based distributions is how the skewness and kurtosis is modelled. In the case of the skewed distributions considered here, and many others, the skewness is modelled explicitly by means of the skewness vector. Moreover, the concentration parameter also allows for the direct modelling of kurtosis. In the case of transformation methods, the skewness and kurtosis are modelled implicitly by means of the transformation vector $\loada$. Therefore, it can be argued that the use of skewed distributions allows for slightly increased interpretability concerning skewness and kurtosis.

%On the other hand, the transformation methods are more parsimonious in terms of the number of free parameters. Specifically, transformation methods have a total of $pG$ additional parameters, when compared to the Gaussian mixture model, from the transformation vector $\loada$. In the case of the skewed distributions considered herein, as well as many others, there are a total of $pG+G$, or in the case of the generalized hyperbolic distribution, $pG+2G$ additional parameters. This is due to the addition of the concentration/index parameter(s). Although not significant for small $G$ and large $n$, this could become significant in the case of larger $G$ and smaller $n$.

\section{Comparison Using Multiple Datasets}
Using the initialization, model selection and convergence criteria outlined previously, we perform a comparison based on multiple real benchmarking datasets.

The first dataset we consider is the iris dataset described previously. Again, this dataset considers 150 observations from three different species of iris, on four variables. The results are summarized in Table \ref{tab:Irisres} along with the skewness, kurtosis with their respective test p-values, and the classification maps. Both the skewness and kurtosis are not significantly different from what would be expected under multivariate normality. Moreover, as discussed previously, and as is well known with this dataset, the first species is well separated from species 2 and 3, and species 2 and three have a fair amount of overlap. The overall performance over these methods is identical in terms of the number of groups chosen, and the classification performance. The likelihood values are very comparable over all methods, and the lowest BIC is obtained by the power transformation; however, given the comparable likelihood values and the additional parameters, this difference is not really substantial.
\begin{table}[!htb]
\centering
{\small
\caption{Results of the skewed models and transformation methods for the Iris dataset.}
{\begin{tabular}{cccccc}
&\multicolumn{2}{c}{Skewed Models}&&\multicolumn{2}{c}{Transformations}\\
\hline
&VG&GH&&Manly&Power\\
\hline
Log-Likelihood&$-307.31$&$-311.04$&&$-308.24$&$-306.81$\\
BIC&810.03&827.51&&801.86&799.01\\
$\mathcal{M}$&39&41&&37&37\\
$\hat{G}$&2&2&&2&2\\
ARI&0.568&0.568&&0.568&0.568\\
Confusion&
$\left(\begin{tabular}{rrr}
     50 &   0 \\ 
        0 &  50 \\ 
        0 &   50\\ 
\end{tabular}\right)$&
$\left(\begin{tabular}{rrr}
     50 &   0 \\ 
        0 &  50 \\ 
        0 &   50\\ 
\end{tabular}\right)$&&
$\left(\begin{tabular}{rrr}
      50 &   0 \\ 
        0 &  50 \\ 
        0 &   50\\ 
\end{tabular}\right)$&
$\left(\begin{tabular}{rrr}
          50 &   0 \\ 
        0 &  50 \\ 
        0 &   50\\ 
     \end{tabular}\right)$\\
\hline
\multicolumn{6}{c}{$G=3$ \quad $p=4$ \quad $n: 50+50+50=150$ }\\[+2pt]
\multicolumn{6}{c}{Skewness: (2.90(0.24), 2.84(0.26), 2.97(0.21))} \\[+2pt]
\multicolumn{6}{c}{Kurtosis: (1.49(0.45), $-$2.03(0.30), $-$0.66(0.74))} \\[+2pt]
 \multicolumn{6}{c}{KDE Map: $\left(\begin{tabular}{rrr}
1.00 & 0.00 & 0.00 \\ 
  0.03 & 0.77 & 0.21 \\ 
  0.00 & 0.17 & 0.83 \\ 
\end{tabular}\right)$ \quad Gaussian Map: $\left(\begin{tabular}{rrr}
1.00 & 0.00 & 0.00 \\ 
  0.00 & 0.98 & 0.02 \\ 
  0.00 & 0.03 & 0.97 \\ 
\end{tabular}\right)$} \\
\hline
\end{tabular}}
\label{tab:Irisres}}
\end{table}

The second dataset considered was the 13 variable wine dataset from the {\sf R} package {\tt rattle} \citep{rattle} which measures 13 chemical properties of 3 three different types of wine. Results are shown in Table \ref{tab:Wineres}. All four methods under fitted the true number of groups. This could be due to one of two reasons. The first is the dimensionality of the data and the fact that we are fitting an unconstrained covariance/scale matrix. However, this could also be due to the significant negative kurtosis for groups 1 and 3. Again, however, there is very little difference in the likelihood and BIC values. Moreover, in terms of classification performance, the results are very similar across methods.
\begin{table}[!htb]
\centering
{\small
\caption{Results of the skewed models and transformation methods for the Wine dataset.}
\begin{tabular}{cccccc}
&\multicolumn{2}{c}{Skewed Models}&&\multicolumn{2}{c}{Transformations}\\
\hline
&VG&GH&&Manly&Power\\
\hline
Log-Likelihood&$-2188.94$&$-2189.30$& &$-2153.58$&$-2144.87$\\
BIC&5605.96&5617.04&&5524.88&5507.45\\
$\mathcal{M}$&237&239&&235&235\\
$\hat{G}$&2&2&&2&2\\
ARI&0.461&0.461&&0.454&0.469\\
Confusion&
$\left(\begin{tabular}{rrr}
     59 &   0 \\ 
   66 &   5 \\ 
    0 &  48 \\
     \end{tabular}\right)$&
$\left(\begin{tabular}{rrr}
     59 &   0 \\ 
   66 &   5 \\ 
    0 &  48 \\ 
     \end{tabular}\right)$&&
$\left(\begin{tabular}{rrr}
         59 & 0\\
      65 & 6\\
       0 &48\\
      \end{tabular}\right)$&
$\left(\begin{tabular}{rrr}
           59 & 0\\
      67 & 4\\
       0 &48\\
               \end{tabular}\right)$\\
\hline
\multicolumn{6}{c}{$G=3$ \quad $p=13$ \quad $n: 59+71+48=178$ }\\ [+2pt]
\multicolumn{6}{c}{Skewness: (47.27(0.37), 57.68(2.35e-11), 50.44(0.96))}\\[+2pt]
\multicolumn{6}{c}{Kurtosis: ($-13.62(8.05\text{e}-3)$, $10.22(0.029)$, $-16.63(3.52\text{e}-3)$)} \\[+2pt]
 \multicolumn{6}{c}{KDE Map: $\left(\begin{tabular}{rrr}
0.91 & 0.08 & 0.00 \\ 
  0.06 & 0.90 & 0.04 \\ 
  0.00 & 0.13 & 0.87 \\ 
\end{tabular}\right)$ \quad Gaussian Map: $\left(\begin{tabular}{rrr}
1.00 & 0.00 & 0.00 \\ 
  0.00 & 1.00 & 0.00 \\ 
  0.00 & 0.00 & 1.00 \\ 
\end{tabular}\right)$} \\
\hline
\end{tabular}
\label{tab:Wineres}}
\end{table}

The next dataset considered was the bankruptcy dataset from the {\sf R} package {\tt MixGHD} \citep{tortora15c} with the results shown in Table \ref{tab:bankruptres}. For all four methods, the BIC chooses one component, and the BIC values are once again comparable. From the skewness and kurtosis measures and from looking at the plot of the data, it is clear that the first component (bankrupt firms) are non-Gaussian and the second group are approximately Gaussian. Moreover, the misclassification maps suggest overlap between these two clusters. This actually displays a general issue in the context of clustering. If a symmetric cluster overlaps with an asymmetric cluster and there is no information about the underlying group structure, then it will be very difficult to determine if there is one cluster or two. Moreover, as seen here, using skewed methods do not help in this scenario, as all four fail to capture the two group solution, and the Gaussian mixture over fits the number of groups. This is an area that should be addressed in future work, as it is quite possible using these methods can result in missing a possibly very important group.

\begin{table}[!htb]
\centering
{\small
\caption{Results of the skewed models and transformation methods for the Bankruptcy dataset.}
\begin{tabular}{cccccc}
&\multicolumn{2}{c}{Skewed Models}&&\multicolumn{2}{c}{Transformations}\\
\hline
&VG&GH&&Manly&Power\\
\hline
Log-Likelihood&$ -114.84$& $-110.85$  &&$-108.94$&$-106.54$\\
BIC& 263.20& 259.40  &&247.21&242.41\\
$\mathcal{M}$& 8&9&&7&7\\
$\hat{G}$&1&1&&1&1\\
ARI&0.000&0.000&&0.000&0.000\\
Confusion&$\left(\begin{tabular}{r}
33 \\ 
   33 \\ 
               \end{tabular}\right)$&
$\left(\begin{tabular}{rr}
33 \\ 
   33 \\ 
             \end{tabular}\right)$&&
$\left(\begin{tabular}{rr}
          33 \\ 
   33 \\ 
          \end{tabular}\right)$&
$\left(\begin{tabular}{rr}
         33 \\ 
   33 \\    
          \end{tabular}\right)$
\\
\hline
\multicolumn{6}{c}{$G=2$ \quad $p=2$ \quad $n: 33+33=66$ }\\[+2pt]
\multicolumn{6}{c}{Skewness: $(15.33(<1\text{e}-16), 0.54(0.56))$} \\ [+2pt]
\multicolumn{6}{c}{ Kurtosis: $(15.42(<1\text{e}-16), -1.43(0.30))$} \\ [+2pt]
\multicolumn{6}{c}{KDE Map: $\left(\begin{tabular}{rr}
0.84 & 0.16 \\ 
  0.12 & 0.88 \\ 
\end{tabular}\right)$ \quad Gaussian Map: $\left(\begin{tabular}{rrr}
0.98 & 0.02 \\ 
  0.02 & 0.98 \\ 
\end{tabular}\right)$} \\
\hline
\end{tabular}
\label{tab:bankruptres}}
\end{table}

The diabetes dataset \citep{fraley12b} considers three measurements on 145 non-obese diabetes patients, with three types of diabetes which were classified as normal, overt and chemical, with results shown in Table \ref{tab:diabetes}. Again, very little difference is seen in the performance of the methods. 
\begin{table}[!htb]
\centering
{\small
\caption{Results of the skewed models and transformation methods for the Diabetes dataset.}
\begin{tabular}{cccccc}
&\multicolumn{2}{c}{Skewed Models}&&\multicolumn{2}{c}{Transformations}\\
\hline
&VG&GH&&Manly&Power\\
\hline
Log-Likelihood&$-178.75$& $-176.62$ & &$-171.68$&$-168.24$\\
BIC&491.88& 497.57 &&467.77&480.80\\
$\mathcal{M}$&27&29&&25&25\\
$\hat{G}$&2&2&&2&2\\
ARI&0.465& 0.450 &&0.465&0.488\\
Confusion&$\left(\begin{tabular}{rr}
32 &   4 \\ 
   76 &   0 \\ 
    2 &  31 \\
                   \end{tabular}\right)$&
$\left(\begin{tabular}{rr}
32 &   4 \\ 
   75 &   1 \\ 
    2 &  31 \\ 
           \end{tabular}\right)$&&
$\left(\begin{tabular}{rr}
     32 & 4\\
      76 & 0\\
       2 &31\\
     \end{tabular}\right)$&
$\left(\begin{tabular}{rr}
    32 & 4\\
      76 & 0\\
       1& 32 \\
          \end{tabular}\right)$\\
\hline
\multicolumn{6}{c}{$G=3$ \quad $p=3$ \quad $n: 36+76+33=145$ }\\[+2pt]
\multicolumn{6}{c}{Skewness: $(9.74(7.09\text{e}-9), 3.45(3.68\text{e}-6), 7.22(1.92\text{e}-5))$}\\ [+2pt]
\multicolumn{6}{c} {Kurtosis: $(8.73(1.72\text{e}-06), 3.22(0.010), 3.28(0.085))$ }\\ [+2pt]
\multicolumn{6}{c}{KDE Map: $\left(\begin{tabular}{rrr}
0.47 & 0.40 & 0.13 \\ 
  0.13 & 0.87 & 0.01 \\ 
  0.09 & 0.25 & 0.66 \\ 
\end{tabular}\right)$ \quad Gaussian Map: $\left(\begin{tabular}{rrr}
0.91 & 0.08 & 0.01 \\ 
  0.02 & 0.98 & 0.00 \\ 
  0.02 & 0.00 & 0.98 \\ 
\end{tabular}\right)$} \\
\hline
\end{tabular}
\label{tab:diabetes}}
\end{table}

We also considered the AIS dataset \citep{sn} which involves 11 measures from 100 female and 102 male athletes. Here we present the results for the three commonly used variables for this dataset, namely the BMI, body fat and lean body mass with the results in Table~\ref{tab:AIS}. Again, very little difference was seen in performance for the generalized hyperbolic and both transformation methods. What is interesting, however, is where these misclassifications lie. Specifically, the variance-gamma misclassifies more men than women whereas the generalized hyperbolic misclassifies more women than men. Moreover, when comparing the transformation methods, equal numbers of men and women are misclassified. 

\begin{table}[!htb]
\centering
{\small
\caption{Results of the skewed models and transformation methods for the AIS dataset.}
\begin{tabular}{cccccc}
&\multicolumn{2}{c}{Skewed Models}&&\multicolumn{2}{c}{Transformations}\\
\hline
&VG&GH&&Manly&Power\\
\hline
Log-Likelihood&$-619.22$& $-618.18$ &&$-620.77$&$-604.21$\\
BIC&1381.76& 1390.29&&1341.12&1347.50\\
$\mathcal{M}$&27&29&&25&25\\
$\hat{G}$&2&2&&2&2\\
ARI&0.847&0.922&&0.922&0.922\\
Confusion&
$\left(\begin{tabular}{rr}
99 & 1 \\ 
   7 &   95 \\ 
      \end{tabular}\right)$&
$\left(\begin{tabular}{rr}
97 & 3\\ 
   1 &   101 \\ 
       \end{tabular}\right)$
&&
$\left(\begin{tabular}{rr}
    98  & 2\\
     2  &100\\    
          \end{tabular}\right)$&
$\left(\begin{tabular}{rr}
    98  & 2\\
     2  &100\\    
          \end{tabular}\right)$\\
\hline
\multicolumn{6}{c}{$G=2$ \quad $p=3$ \quad $n: 100+102=202$ }\\[+2pt]
\multicolumn{6}{c}{Skewness: $(2.54(6.53\text{e}-6), 5.66(3.33\text{e}-16))$} \\ [+2pt]
\multicolumn{6}{c}{Kurtosis: $(1.69(0.12), 7.97(2.00\text{e}-13)) $}\\ [+2pt]
\multicolumn{6}{c}{KDE Map: $\left(\begin{tabular}{rr}
0.89 & 0.11 \\ 
  0.09 & 0.91 \\
\end{tabular}\right)$ \quad Gaussian Map: $\left(\begin{tabular}{rrr}
0.98 & 0.02 \\ 
  0.02 & 0.98 \\ 
\end{tabular}\right)$} \\
\hline
\end{tabular}
\label{tab:AIS}}
\end{table}

The last dataset considered herein was the famous crabs dataset, \cite{venables02}. This considers two species of crab (blue and orange) and males and females within each species. The first group are the blue males, the second the blue females, the third the orange males, and the fourth the orange females. The results are shown in Table \ref{tab:crabs}. What is very interesting, but not entirely clear from the overlap, skewness, and kurtosis from the four groups is that the skewed distribution methods separate the species perfectly, whereas the transformation methods discriminate based on gender. As all methods were run to convergence on many different initialization values, and the initializations were the same for each method, it is unlikely that it is due to the initialization. However, Table \ref{tab:crabsskew} shows the skewness and kurtosis values and their respective p-values based on sex and species, and it appears that the transformation methods found one skewed component and a symmetric component, whereas the skewed distribution methods found two skewed components. This might suggest that the transformation methods might be slightly more likely to find symmetric components than the skewed distributions. 

\begin{table}[!htb]
\centering
{\small
\caption{Results of the skewed models and transformation methods for the Crabs dataset.}
\begin{tabular}{cccccc}
&\multicolumn{2}{c}{Skewed Models}&\multicolumn{2}{c}{Transformations}\\
\hline
&VG&GH&Manly&Power\\
\hline
Log-Likelihood&165.25&162.12 &144.68&144.62\\
BIC&$-49.69$&$-32.83$&$-19.16$&$2.17$\\
$\mathcal{M}$&53&55&51&51\\
$\hat{G}$&2&2&2&2\\
ARI&0.496&0.496&0.374&0.374\\
Confusion&
$\left(\begin{tabular}{rrr}
     50 &   0 \\ 
   50 &   0 \\ 
    0 &  50 \\ 
    0 &  50 \\
          \end{tabular}\right)$&
$\left(\begin{tabular}{rrr}
     50 &   0 \\ 
   50 &   0 \\ 
    0 &  50 \\ 
    0 &  50 \\ 
     \end{tabular}\right)$&
$\left(\begin{tabular}{rrr}
         47  &3\\
       6 &44\\
      50 & 0\\
       4 &46\\
               \end{tabular}\right)$&
$\left(\begin{tabular}{rrr}
                   47 & 3\\
       6 &44\\
      50 & 0\\
       4 &46\\
           \end{tabular}\right)$\\
\hline
\multicolumn{6}{c}{$G=4$ \quad $p=5$ \quad $n: 50+50+50+50=200$ }\\[+2pt]
\multicolumn{6}{c}{Skewness: $(4.71(0.28), 5.01(0.20), 2.82(0.93),  5.10(0.18))$} \\ [+2pt]
\multicolumn{6}{c}{Kurtosis: $(-1.26(0.60), -0.80(0.73), -3.17(0.18), -0.18(0.94))$} \\[+2pt]
 \multicolumn{6}{c}{KDE Map: $\left(\begin{tabular}{rrrr}
0.33 & 0.25 & 0.29 & 0.13 \\ 
  0.20 & 0.38 & 0.20 & 0.22 \\ 
  0.26 & 0.18 & 0.37 & 0.19 \\ 
  0.15 & 0.31 & 0.22 & 0.32 \\ 
\end{tabular}\right)$ \quad Gaussian Map: $\left(\begin{tabular}{rrrr}
0.95 & 0.05 & 0.00 & 0.00 \\ 
  0.04 & 0.96 & 0.00 & 0.00 \\ 
  0.00 & 0.00 & 0.99 & 0.01 \\ 
  0.00 & 0.00 & 0.01 & 0.98 \\ 
\end{tabular}\right)$} \\
\hline
\end{tabular}
\label{tab:crabs}}
\end{table}

\begin{table}
\centering
\caption{Skewness and kurtosis for the crabs dataset based on sex and species separately.}
\begin{tabular}{ccc}
& Skewness (p-value) & Kurtosis (p-value)\\
\hline
Males & $2.7 (0.12)$ & $-2.38 (0.15)$\\
Females & $3.64 (0.0046)$ & $-0.47 (0.78)$\\
\hline
Blue & $4.9 (1.30\text{e}-5)$ & $0.87 (0.6)$\\
Orange & $4.01 (9.50\text{e}-4)$ & $0.37 (0.83)$\\
\hline
\end{tabular}
\label{tab:crabsskew}
\end{table}

\section{Discussion}
From the analyses performed on a variety of datasets and the extensive number and type of initializations performed, it appears that no one method consistently outperforms the others, and usually the performance is very similar if not identical. Moreover, it does not appear that skewness, kurtosis and cluster overlap completely determines the relative performance of these methods. This is not to say, however, that there are no differences between the two methods. As seen with the crabs, the skewed models discriminate based on species, and the transformation methods on sex. Although not shown herein, the skewed models also discriminate based on sex when using k-means initializations, but when run using the more flexible initialization methods, found the species structure. The transformation methods, on the other hand, always found the sex structure. In terms of actual properties of the methods, transformation methods are more parsimonious, controlling for the number of groups, due to the lack of a concentration (or index) parameter.
 Therefore, it may not be a question of when one method might be preferable to another, but rather why one method might be preferable to another in the context of the analysis in question.

\end{document}